\documentclass[compsoc, conference, a4paper, 10pt, times]{IEEEtran}
\usepackage{cite}
\usepackage{amsmath,amssymb,amsfonts}
\usepackage{graphicx}
\usepackage{textcomp}
\usepackage{xcolor}
\usepackage{booktabs}
\usepackage[hidelinks]{hyperref}
\usepackage[inline]{enumitem}
\usepackage{tikz-cd}
\usepackage{xurl}

\usetikzlibrary{fit,shapes.geometric}
\tikzcdset{every arrow/.append style = -latex}

\usepackage{bm}
\usepackage{algorithm, algpseudocodex}
\def\BibTeX{{\rm B\kern-.05em{\sc i\kern-.025em b}\kern-.08em
    T\kern-.1667em\lower.7ex\hbox{E}\kern-.125emX}}

\usepackage[capitalize, nameinlink, noabbrev]{cleveref}
\newtheorem{definition}{Definition}

\floatname{algorithm}{Protocol}
\crefname{algorithm}{Protocol}{Protocols}

\newcommand{\F}{\mathcal{F}}

\begin{document}

%\title{Conference Paper Title*\\

%proposed title by Meiko
\title{Identification of Compositional Risks\\in Data Protection Impact Assessments and Beyond}

\author{
\IEEEauthorblockN{Henrik Gra\ss hoff, Meiko Jensen}
\IEEEauthorblockA{
\textit{Karlstad University}\\
Karlstad, Sweden \\
\{henrik.grasshoff, meiko.jensen\}@kau.se}
\and 
\IEEEauthorblockN{Malte Hansen, Nils Gruschka}
\IEEEauthorblockA{
\textit{Department of Informatics, University of Oslo} \\
Oslo, Norway \\
\{maltehan, nilsgrus\}@ifi.uio.no}
}

\maketitle

\begin{abstract}
When personal data is processed in a distributed manner by cooperating service providers, privacy risks may emerge solely from the choice of data processors included in the composition. For instance, different data processors may unknowingly rely on the same cloud provider, allowing for unintended linkability of personal data at that very provider. As such \emph{compositional risks} to privacy are beyond the scope of each individual risk assessment, they are likely to be overseen when performing a data protection impact assessment.

In this paper, we propose a novel protocol to detect and manage such compositional risks to privacy. Following an initial problem definition and requirements elicitation, we elaborate how our protocol identifies candidates for compositional risks and how this information may be used to improve the results of a data protection impact assessment over service compositions including multiple data processors.
\end{abstract}


\section{Introduction and Motivation}

According to Art.~35 of the General Data Protection Regulation (GDPR)~\cite{GDPR}, every data processing system deployed in the European Union needs to undergo at least the threshold analysis for a Data Protection Impact Assessments (DPIAs).
A key element in such a DPIA is the detection of risks towards the rights and freedoms of data subjects caused by the processing of their personal data, which need to be documented---along with their probability and damage assessments, as well as potential mitigation measures applied---as part of the DPIA result documentation (cf.~\cite{wp29_dpia_guidelines,martin2020data}).

In real-world contexts, the set of risks existing in data processing instances may span over multiple data processors, which may or may not collaborate in the joint DPIA process led by the responsible data controller(s)\footnote{According to Art.~26 GDPR, several different organisations may act as joint controllers towards a data processing instance. However, for the sake of readability, in this paper, we utilise the singular form to reflect the role of the actor(s) who lead(s) the DPIA process. This can be a joint effort by all (or a subset of) data controllers, but can also reside with one of the data controllers only.}. Moreover, in addition to the individual risks arising at each data processor, more complex types of risks may result from the \textit{composition} of data processors rather than from the individual processing at each single entity.

In this paper, we discuss the nature and specifics of such compositional risks to privacy as found in modern composed data processing systems, and we propose a protocol to detect such compositional risks as part of a DPIA process. Our detection protocol allows for the identification of data processor constellations that may result in compositional risks, and it takes into account the trust model and needs for business confidentiality among the actors involved in a composed data processing instance that spans across arbitrary large service compositions.

The paper is organised as follows. First, we provide some background on DPIAs and their regulatory requirements (\cref{secBackground}). In \cref{secProblemStatement}, we elaborate the problem statement, illustrate the specific properties of compositional risks in modern service compositions, and derive a set of requirements towards our proposed solution approach. The abstracted graph model and distributed protocol for identification of potential compositional risks are then presented in \cref{secAlgorithm}. \cref{sec: handling,secDiscussion} give indications on how to proceed with detected potential risks and discuss the protocol's implementation. The paper concludes with future work in \cref{secConclusion}.

\section{Background}

%\subsection{Data Protection Impact Assessments}
\label{secBackground}
The GDPR introduced DPIAs as a mandatory requirement for data controllers before the commissioning of new information systems or processing activities that are likely to lead to high risks to the personal rights and freedoms of individuals~\cite[Art.~35~(1)]{GDPR}. As a formalised process of a privacy impact assessment, a DPIA, among other things, must contain a systematic description of the processing operations, an assessment of the risks to the individuals resulting from these operations, as well as a description of measures on how to address these risks~\cite[Art.~35~(7)]{GDPR}.

In this respect, several different methodologies and frameworks for Data Protection Impact Assessments have been proposed by data protection authorities as well as in the scientific literature (cf.~\cite{DSK_DPIA,CNIL_DPIA, martin2020data}). Although not explicitly required by the GDPR, some DPIAs have also been made publicly available~\cite{EC_smart_meter_DPIA,NL_DPIA}, showcasing the different approaches used for risk identification and assessment in practice. Beyond sheer enumeration of identified risks, the documentation of respective risk mitigation measures implemented throughout the processing, such as utilised privacy-enhancing technologies, are an essential part of each DPIA assessment as well.

With data sharing becoming increasingly more prominent, processing operations will often-times include more entities, e.g. one or multiple data processors, rather than just a single data controller. To give an example, the specific constellations of actors in a Common European Data Space was identified as a risk source by ENISA~\cite{enisa_data_spaces}. In order to fulfil the aforementioned requirements of a DPIA, these data flows and external risks must be included in the DPIA process. The European Data Protection Board clarifies that the data controller is responsible for the data processors in these cases and must therefore cover their involvement in the DPIA as well, asking for assistance where necessary~\cite{wp29_dpia_guidelines}. The Spanish data protection authority AEPD further backs this by stating that the DPIA for large scale data processing scenarios with multiple actors should consider the approach to such collaborations~\cite[p.~69]{aepd_dataspaces}.

The problem considered in this paper is related to the field of cloud computing, where the detection and mitigation of security risks have been objects of extensive research.
With reference to the processing of sensitive data, malicious coalitions of two or more service providers is regarded a major threat to the confidentiality of these data. Existing methodologies primarily aim at preventing breaches by either partitioning business processes into smaller pieces before or during deployment (cf.~\cite{leite2014deploying,nacer2017metric,rekik2016comprehensive}) or by detecting and managing the collusion of providers~\cite{ahmed2024malicious}. In this paper, we address a number of limitations of these methods and propose a protocol that
\begin{enumerate*}
    \item considers the mere coincidence of sub-providers as a potential risk,
    \item is applicable after the service composition has been established, and
    \item allows to detect these constellations in a distributed manner without the need to reveal business relationships.
\end{enumerate*}

\section{Problem Statement}
\label{secProblemStatement}

As discussed above, DPIAs require the identification and assessment of risks to the rights and freedoms of individuals caused by processing of their personal data~\cite{GDPR}. It is the obligation of the data controller(s) to perform and manage the assessment and to document the results. In practice, as was noticed by \cite{friedewald2021data}, performing a DPIA risk assessment requires not just work efforts of the data controller, but must also be supported from the data processors involved in a data processing system. Specific risks may reside only with some data processors, yet, these need to be documented as part of the DPIA process led by the data controller. For instance, the risk of potential leakage of personal data to unauthorised third parties manifests differently within different data processor organisations. Hence, it becomes necessary to approach all data processors that are part of a service composition and collect their specific inputs to the list of risks identified to be assessed as part of the DPIA process.

Beyond the sheer concatenation or aggregation of the individual lists of risk sources at each individual data processor, however, the selection of data processors and their sub-processors itself may cause new types of risks. Such \textit{compositional data protection risks} do not originate from the specific modalities of data processing at a single data processor organisation, but stem from the composition of data processors, sub-processors, sub-sub-processors, etc. involved in a compositional data processing instance.

\subsection{Compositional Risk Examples}
\label{secExample}

\begin{figure*}
    \centering
    \includegraphics[width=0.66\textwidth]{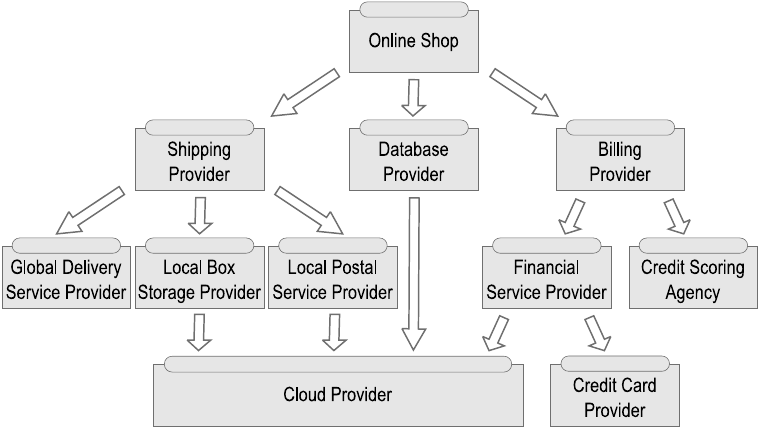}
    \caption{Example scenario of a service composition}
    \label{fig:servicecomposition}
\end{figure*}

\cref{fig:servicecomposition} shows a typical example of a composed service that follows the service-oriented architecture paradigm. As can be seen, the \texttt{Online Shop} (as the data controller in this scenario) utilises a set of data processors to realise the implementation of all required elements of the data processing. Each such data processor has its own set of sub-processors that provide required functionality towards the specific services implemented by the data processor. This list of sub-processors may be empty (a \emph{sink} in the data processing graph), or it may consist of a list of subsequent sub-processors. For instance, the set of sub-processors of the \texttt{Billing Provider} is the set \texttt{\{Financial Services Provider, Credit Scoring Agency\}}, whereas the sub-processor list of the \texttt{Credit Scoring Agency} is empty.

It is very common for many of today's service implementations that several data processors rely on the services of a \texttt{Cloud Provider}, e.g. for data storage or for provisioning of computational resources. Given today's distribution of market shares in the cloud market~\cite{feil2024market}, it is rather likely that two or more different data processors decide to utilise the very same cloud provider as sub-processor for their implementations (as illustrated in \cref{fig:servicecomposition}). In this case, the \texttt{Cloud Provider} hence acts as sub*-processor\footnote{In the remainder of this paper, the term \emph{sub*-processor} refers to an actor who is a sub-processor at \emph{some} level of depth within the service composition.} at different places in the data processing instance covered by the risk assessment.

If we now assume that the customer records (including name, address, etc.) of the data controller's customers are forwarded to and stored at several different data processors (e.g. as a billing address with the \texttt{Billing Provider} and as a shipping address with the \texttt{Shipping Provider}), the very same data may end up in different accounts of the very same \texttt{Cloud Provider}. Hence, it becomes possible for said \texttt{Cloud Provider} to identify the address in multiple datasets and link the different contexts across its customers. This way, the \texttt{Cloud Provider} may for instance be able to link the credit scores of a data subject to its shopping lists (and habits), allowing for potential profiling of \texttt{Online Shop} customers at the \texttt{Cloud Provider}. This in itself may already enable the linking of separated datasets, potentially violating the unlinkability protection goal of privacy~\cite{hansen2015protection}, and hence causing novel linkability risks to consider within the DPIA process.

\begin{figure}[b]
    \centering
    \includegraphics[height=3.5cm]{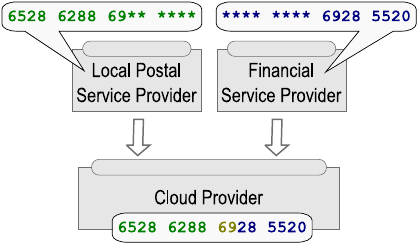}
    \caption{Mutually eliminating credit card masking}
    \label{fig:ccmasking}
\end{figure}

\begin{figure}[b]
    \centering
    \includegraphics[height=3.3cm]{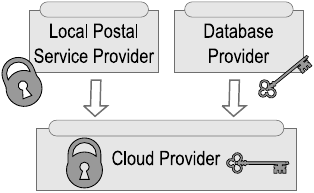}
    \caption{Broken distribution of ciphertext and key}
    \label{fig:keyAndCiphertext}
\end{figure}

\subsubsection{Credit Card Number Masking}
\label{exampleCCNumber}

Similarly to the previous example, a new compositional risk may stem from the application of privacy-enhancing technologies (PETs) at each data processor.
If the \texttt{Local Postal Service Provider} stores the customer's credit card number at the \texttt{Cloud Provider}, but replaces the \textit{last} six digits with a ${\ast}$ character, while the \texttt{Financial Service Provider} does the same, but masks out the \textit{first} eight digits of said number, the \texttt{Cloud Provider} can easily reconstruct the complete credit card number by linking the two parts from the two different contexts. The overlap of non-masked digits would not even be relevant here, as the context of belonging to the very same customer dataset would easily allow for linkability of the components without it. A very similar real-world incident took place in 2011 (cf.~\cite{schneier2011})\footnote{The industry standard for processing credit card numbers imposes restrictions on the way credit card numbers are masked~(cf. Requirement~3.3 in PCI-DSS~\cite{PCIDSS}). However, those restrictions do not completely eliminate the risk described here, as was discussed in~\cite{schneier2011}.}.
As can be seen, the privacy-enhancing technology of masking that was applied to the credit card number independently at the \texttt{Local Postal Service Provider} and at the \texttt{Financial Service Provider} respectively is trivially broken at the \texttt{Cloud Provider}, based on the composition of actors (see \cref{fig:ccmasking}).

It is important to note that even the \texttt{Cloud Provider} is typically not aware of this compositional risk. From its perspective, both masked credit card number inputs come from different cloud customers and datasets, hence, it is not obvious to the \texttt{Cloud Provider} that these may result from the same service composition and data processing instance. However, once the link between these two contexts becomes identifiable, an unmasking attempt is trivial to perform at the \texttt{Cloud Provider}.

\subsubsection{Key and Ciphertext}
\label{exampleKeyAndCiphertext}

Another example of a compositional risk can be identified when data is stored in encrypted form, so as to hide its contents from subsequent data processors (such as the \texttt{Cloud Provider}). As shown in \cref{fig:keyAndCiphertext}, assume the \texttt{Online Shop} decides to encrypt (parts of) the customer's data records in order to protect privacy and forwards only the resulting ciphertext to its data processors while storing the secret key in its own database only (at the \texttt{Database Provider}, which itself utilises the \texttt{Cloud Provider}). Then, subsequently, the \texttt{Local Postal Service Provider} in charge for a customer's order may store said ciphertext (which was received via the \texttt{Shipping Provider}) in its own cloud storage account (at the very same \texttt{Cloud Provider}). If the \texttt{Cloud Provider} is now able to link the two data records received from the \texttt{Local Postal Service Provider} and from the \texttt{Database Provider}, it is also able to utilise the secret key to decrypt the data records---breaking the assumed protection it provides.
Again, this risk emerges solely from the fact that two different data processors utilise the very same \texttt{Cloud Provider} as a sub-processor. Even worse, each individual data processor assumes it has little (or no) residual risk towards the customer data (as it is either a ciphertext or just a secret key, not the data itself). Hence, when assessing the risk of data record disclosure, both the \texttt{Database Provider} and the \texttt{Local Postal Service Provider} may assume that risk to be of low or even zero probability due to the application of encryption, whereas, factually, the risk of decryption and leakage at the \texttt{Cloud Provider} (by means of the \texttt{Cloud Provider} itself or of another entity that manages to hack the \texttt{Cloud Provider} or its user accounts) is substantial.

\subsection{Compositional Risk Identification}
\label{subsec: compo risk ident}

It is important to understand that each of the risk examples described above is solely based on the composition of data processors. For each individual data processor, the risk might not be of relevance---or not even be visible due to the lack of overview over the service composition as a whole. Only when assessing the full composition, these potentially problematic constellations may become visible. This is a key aspect of such compositional risks.

How can one then reasonably identify the presence of compositional risks in a composed service implementation with several data processors and sub*-processors?

A key observation with respect to such compositional risks is that they can only manifest if different data processors utilise the same sub-processor for their data processing implementation. 
In the example of \cref{fig:servicecomposition}, this was the case for the same \texttt{Cloud Provider} being utilised by several different data processors.
More generically, we assume that all constellations involving the same sub-processor by more than one data processor potentially raise compositional risks, and identification of those risks (or validation of their non-existence) can only be done when such constellations are properly identified early, e.g. in the context of a compositional DPIA risk assessment.

\subsection{Compositional Risk Assessment versus\\Service Composition Confidentiality}
\label{subsec: service confidentiality}

If the data controller in the example scenario of \cref{secExample} was aware of all data processors and sub*-processors involved in the whole data processing instance, detection of potential risk constellations as discussed above would be trivial. As the data controller could construct the whole processing tree, it could trivially detect critical constellations and investigate each of those specifically for potential compositional risks manifesting.

However, not all data processors share their list of sub*-processors with each of their service customer organisations freely. Keeping the identity of their suppliers confidential may often be seen as a necessary approach to safeguard business secrets (cf. e.g.~\cite{Jensen2013}) or may reduce the threat of being excluded from a service implementation---and hence loosing customers. This sort of business partner confidentiality often conflicts with the intentions behind a DPIA as the data controller is liable for the whole data processing instance and should thus be able to know and control each aspect of the data processing if necessary. On the other hand, data processors like cloud providers often do not want their customers to learn the risks involved in utilising their services for processing and especially not those induced by their respective sub*-processors.

Hence, when it comes to the identification of compositional risks, the bird's-eye approach of collecting the whole composition picture at the data controller to locally check for potential compositional risks is not always feasible. A valid identification approach for compositional risks has to consider the protection of business partner identities and service relationships among the sub*-processors instead as far as possible in order to become applicable in real-world scenarios.

Especially each potential source of compositional risks, such as the \texttt{Cloud Provider} in \cref{fig:servicecomposition}, may have an implicit motivation to hide the fact that it could be perceived as a compositional risk threat as one potential and trivial risk elimination strategy might consists in choosing a different cloud storage provider at each involved sub*-processor. Thus, if identified as the potential cause of a compositional risk, such a sub*-processor may decide to hide the fact that it is involved in two or more different sub-processes of the same data processing instance---and even worse, may decide to utilise this knowledge to secretly search for linkability within the data it receives from the involved sub*-processors to actively exploit the compositional risk itself.

\subsection{Requirements Towards a Protocol}
\label{subsec: requirements}

Based on the observations above, when developing a protocol that allows for detection of critical constellations towards a compositional risk in a service composition, it becomes necessary to optimise towards the following list of requirements:
\begin{enumerate}[label=\texttt{R\arabic*)}]
    \item After execution of the protocol, \emph{authorised} entities in the composition graph learn whether and how many potential candidates for compositional risks exist. \label{item: correctness}
    \item A sub*-processor that potentially causes a compositional risk does not learn about this fact from the protocol. \label{item: hiding}
    \item Any entity learns only as little as possible about the service composition graph outside of its inherent view (i.e. its parent and children nodes). \label{item: topology}
    \item The execution of the protocol must be computationally feasible. \label{item: last}
\end{enumerate}

In order to abstract from the specifics of DPIA constellations in real-world settings, we reduced the problem scenario to a generic model for service compositions that resembles a directed graph of service interactions. In this graph, each node represents a sub*-processor involved in the respective data processing instance, and each directed edge reflects a service interaction between a data processor and one of its sub-processors. Based on this model, we derive a protocol that outputs identifiers of potential candidates for compositional risks in the data processing instance. The subsequent means to validate or negate the existence of actual compositional risks in these constellations, however, is considered out of scope of this work as it largely depends on specifics of the respective implementations at the involved actors.

\section{Detection Protocol}
\label{secAlgorithm}

In this section, we propose a protocol which allows for the detection of potential compositional risks with respect to the requirements \ref{item: correctness}--\ref{item: last} above.
We begin by transferring the described scenarios into a more abstract graph-theoretical setting before we define the protocol, derive its properties, and illustrate an example execution.

Generally, the service interactions of a particular data procession instance form a finite, connected, directed graph as in \cref{fig:servicecomposition}.
Although its precise topology is unknown to all parties for the reasons described in \cref{subsec: service confidentiality}, every such graph admits a \emph{source} by definition, i.e. a node $S$ without in-going edges (which resembles the controller).
By the above considerations, in such a graph, compositional risks can occur at any sub*-processor which potentially serves multiple preceding data processors.
This is captured by the following notion:

\begin{definition}
    \label{definition: focal point}
    Let $v,w$ be two nodes in a directed graph $G$.
    We say that $w$ is a \emph{focal point} in $G$ if $w$ has at least two in-going edges.
    In this case, $v$ is a \emph{cusp} for $w$ if it admits at least two distinct paths\footnote{For the sake of unambiguity, we clarify that a \emph{path} in a graph should be understood to be a walk consisting of edges whose endpoints are pairwise different. In particular, a cycle is also considered a path.} to $w$.
\end{definition}
By this definition, the service composition in \cref{fig:servicecomposition} contains \texttt{Cloud Provider} as a focal point with \texttt{Shipping Provider} and \texttt{Online Shop} as its cusps.
The graph in \cref{fig: example graph} admits the focal points $E, G, J$ with suitable cusps $S, A$ (for all points), $B$ (for $E, G,$), $D$ (for $G$), and $J$ (for itself).

\subsection{Cryptographic Prerequisites}
\label{subsec: prerequisites}

Our protocol uses two cryptographic ingredients in the form of Bloom filters and blindable pseudonym functions. In the following, we briefly explain the use and purpose of these schemes up to the necessary level of detail.

\subsubsection{Bloom Filters}

A Bloom filter is a probabilistic data structure $\F$ comprised of an $m$-bit array initialised with zeroes~\cite{bloom1970space}. The filter comes with $k$ different hash functions which map elements to one of the array positions $1, \dotsc, m$ with equal probability and independently of each other. An element can be added to the filter by computing its $k$ hash values and setting the array bits at these positions to $1$. To test whether an element has been added to $\F$, one feeds it into the $k$ hash functions and checks the corresponding array positions; if all of these are ones, then its is \emph{possibly} in the filter, whereas a single zero ensures that it has not been added before. Bloom filters thus allow false positives with a well-known probability which depends on $m, k$, and the number of added items~\cite{gopinathan2020certifying}. Based on Bloom's very simple concept, a large body of research has since then focused on improving its false positive rate and implementation, see~\cite{luo2018optimizing}.
The purpose of the Bloom filter in our context is to guarantee termination when cycles such as the $J$-$K$-$L$-cycle in \cref{fig: example graph} are present.
This comes at the cost of guaranteed correctness (as the recursive calls may terminate too early, cf.~\cref{subsubsec: threat model}).
However, we emphasise that one can choose a specific Bloom filter implementation among the many existing to realise any desired correctness probability---the most naive approach would be to simply choose a filter size $m$ which is large enough for all real-world scenarios.

\subsubsection{Blindable Pseudonyms}

A blindable pseudonym function is a function $p$ to calculate a pseudo-random pseudonym $p(x,k)$ when fed with a salt $x$ and a secret $k$ (e.g. a private cryptographic key) which admits the following blinding property:
\smallskip
\begin{quote}
    The salt $x$ can be altered by a blinding factor $r$ to obtain a different salt $y$ such that it is possible to calculate $p(x, k)$ solely by knowing $p, p(y, k)$, and $r$ without learning anything about $k$.
\end{quote}
\smallskip
Moreover, we assume that collisions occur with practically negligible probability if salt or key remains fixed.
The purpose of blindable pseudonyms is two-fold: First, it prevents nodes from linking two inquiries which originate from the same protocol execution, and second, it prevents node from linking received pseudonyms from different protocol executions.
Blindable pseudonym functions can be implemented by adopting blind signature schemes.

% DO NOT DELETE THIS PARAGRAPH:
%The above property resembles a standard property of blind signature schemes. Thus, as a illustrative and preliminary implementation, one may use the following ElGamal-inspired construction: Let $p$ be a large prime number and $g$ a generator of the multiplicative group $G = (\mathbb{Z}/p\mathbb{Z})^*$ of the field with $p$ elements. Similar to ElGamal, this construction will be based on the hardness of the Discrete Logarithm Problem in $G$. Every party generates a secret key $k$ randomly from $\{2, \dotsc, p-2\}$. A salt $x$ is blinded to $y = x^r$ by a blinding factor $r$ with $\gcd(r,p-1) = 1$. Pseudonyms are calculated as $p(x, k) = x^k$, so that the equation $p(x,k) = p(y,k)^{r^{-1}}$ holds and can be used to unblind $p(x,k)$ from $p(y,k)$ and $r$. This preserves the secrecy of the secret keys $k$, as deriving $k$ from $p(y,k) = y^k$ and $y$ would require an adversary to find discrete logarithms in the group $G$.

\subsection{The Protocol}

The protocol conceptually works by aggregating pseudonyms at each node such that pseudonyms of focal points appear at least twice with an assessable and overwhelming probability. Before the execution, all parties are expected to have agreed upon a blindable pseudonym function $p$ and an implementation of Bloom filters, e.g. via a standardisation.
The source $S$ generates a random number $x$ (``salt'') and instantiates an empty Bloom filter $\F$ and it passes the pair $(x, \F)$ to each of its children.

The subsequent \emph{aggregation phase} then follows a recursive depth-first search approach.
Each node $v$, after receiving a pair $(y, \F)$ from a parent, computes its pseudonym $p(y, k_v)$ for the salt $y$ and its $k_v$, inserts $k_v$ into $\F$, blinds the salt, passes the resulting pair to each of its children, unblinds the pseudonyms they return and sends the combined multiset of unblinded pseudonyms back to the parent node which queried it. The termination condition for this recursion is that either $v$ is a sink in the graph or ``Is $k_v$ in $\F$?'' returns ``True'' (with an assessable probability of a false positive, see \cref{subsec: prerequisites}). In this case, $v$ returns $p(y, k_v)$ to the parent node without querying any children.

The focal point detection can be performed at any node $v$ once it has received the multisets from all of its children.
By checking for duplicates, $v$ can deduce that a focal point exists without learning about their identity or their location in the graph.
This comes with a caveat of failure which we discuss in \cref{subsubsec: threat model} below.

\begin{algorithm}
    \caption{Probabilistic Focal Point Detection}
    \begin{algorithmic}[1]
        \Ensure Directed finite graph $G$ with unique source $S$
        \Require Blindable pseudonym function $p$ as above
        
        % Ugly hardcoding, I know
        \medskip
        \hspace{-1.63cm} \textbf{Initialisation:}
		\State $S$ generates random salt $x$ and empty Bloom filter $\F$
        \State Every party $v$ holds a secret $k_v$
        \State $S$ propagates $(x, \F)$ to its children
        
        \medskip
        \hspace{-.72cm} \textbf{Aggregation phase:}
        
        \hspace{-.72cm} When a node $v$ receives a pair $(y, \F)$ from a parent $u$:
        \State Initialise multiset $P = \{p(y, k_v)\}$
        \If {$k_v$ is not an element of $\F$} \label{line: Bloom check}
            \State Add $k_v$ to $\F$
            \State Blind $y$ by a random noise to obtain $z$ \label{line: blind}
            \For {each child $w$ of $v$}
                \State Send $(z, \F)$ to $w$
                \State $Q \gets \text{multiset received from $w$}$
                \For {each pseudonym $p(z, \text{--})$ in $Q$}
                    \State Calculate $p(y, \text{--})$ from $p(z, \text{--})$ and add this to $P$
                \EndFor
            \EndFor
        \EndIf
        \State Return $P$ to $u$

        \medskip
        \hspace{-.72cm} \textbf{How a node $\bm{v}$ detects focal points:}
        
        \hspace{-.72cm} After $v$ received all multisets from all of its children:
        \If {the union of the multisets contains a duplicate or \\ \hspace*{.24cm} $v$'s pseudonym}
            \State Potential focal point found!
        \EndIf
    \end{algorithmic}
    \label{alg:protocol}
\end{algorithm}

\subsection{Analysis}

In the following, we derive the protocol's properties when executed by honest-but-curious participants and illustrate this with an example.

\subsubsection{Threat Model and Parameter Choice}
\label{subsubsec: threat model}

We assume that all nodes behave in an \emph{honest-but-curious} manner. Every party follows the protocol as specified, but tries to exploit any knowledge leakage for its own benefit.

Apart from that, \cref{alg:protocol} is a probabilistic protocol which admits a certain likelihood of failure.
First, it is theoretically possible that two different nodes generate the same pseudonym by chance, leading a cusp into the false conclusion that a focal point exists.
Second, a Bloom filter check in \cref{line: Bloom check} can return a false positive, which would result in a premature stopping of the recursion and ultimately lead to potentially undetected focal points.
Both concerns can be addressed by an appropriate choice of parameters.
Pseudonyms need to be drawn from a sufficiently big set, and the large amount of existing Bloom filter optimisations~\cite{luo2018optimizing} can be utilised to reduce the chance of failure to an acceptable minimum.
We defer the discussion of choosing these parameters to a future work as we propose they should depend on the expected graph size (which could be derived from market analyses) and balance correctness with the aspects of computational feasibility.
Against this background, we assume in the following discussion that no pseudonyms from different nodes coincides and that Bloom filter checks are correct.

\subsubsection{Properties}
\label{subsubsec: properties}

When executed as specified, \cref{alg:protocol} admits the following properties:
\begin{enumerate}
    \item The protocol terminates. \label{prop: termination}
    \item Every node learns either whether it is no cusp at all or it learns an upper bound for the number of its focal points. \label{prop: learning}
    \item A node cannot distinguish whether two inquiries from two parent nodes belong to the same protocol execution or not. \label{prop: unlinkability}
    \item A focal point which is not its own cusp does not learn from the protocol that it is a focal point. \label{prop: hiding}
    %\item If the graph admits no focal points, its source learns about this fact. \label{prop: assurance}
\end{enumerate}
Property~\ref{prop: termination} is immediate from the finiteness of the graph and the protocol's recursive nature: The only barrier to the protocol's termination is an infinite number of recursion calls---which is impossible due to the Bloom filter check.
Property~\ref{prop: learning} states that the protocol regards every cusp as an authorised entity in the sense of~\ref{item: correctness} in \cref{subsec: requirements}.
Moreover, it can be used to test for the absence of focal points, which may be of particular interest for the graph source to know.
Property~\ref{prop: unlinkability} is achieved by blinding the salt in \cref{line: blind} before propagation:
From a node's perspective, all received queries thus appear with a random salt and are unlinkable.\footnote{This does evidently not exclude statistical or timing attacks one can perform to link queries, a problem whose prevention is outside of the scope of this work.}
From this, we can conclude Property~\ref{prop: hiding} which realises~\ref{item: hiding}.
The only case in which a focal point \emph{does} learns that is a focal point is when it is part of a cycle, e.g. the node $J$ in the $J$-$K$-$L$-cycle in \cref{fig: example graph}.
It is up to a choice in this situation to either allow the node to learn about its focal-point property or to keep this information hidden---\cref{alg:protocol} as specified favours the detection over the concealing here.

\subsubsection{Example}
\label{subsubsec: example}

\begin{figure}[h!]
    \centering
    \begin{tikzpicture}
        % Nodes
        \node (S) at (0,6) {$S$};
        \node (A) at (0,5) {$A$};
        \node (B) at (-1.5,4) {$B$};
        \node (I) at (0,4) {$I$};
        \node (J) at (1.5,4) {$J$};
        \node (C) at (-2,3) {$C$};
        \node (D) at (-1,3) {$D$};
        \node (K) at (1.5,3) {$K$};
        \node (E) at (-2,2) {$E$};
        \node (F) at (-1,2) {$F$};
        \node (L) at (1.5,2) {$L$};
        \node (G) at (-1.5,1) {$G$};
        \node (H) at (-1.5,0) {$H$};
       
        % Arrows
        \draw[-latex] (S) -- (A);
        \draw[-latex] (A) -- (B);
        \draw[-latex] (A) -- (I);
        \draw[-latex] (A) -- (J);
        \draw[-latex] (B) -- (C);
        \draw[-latex] (B) -- (D);
        \draw[-latex] (J) -- (K);
        \draw[-latex] (C) -- (E);
        \draw[-latex] (D) -- (E);
        \draw[-latex] (D) -- (F);
        \draw[-latex] (K) -- (L);
        \draw[-latex] (E) -- (G);
        \draw[-latex] (F) -- (G);
        \draw[-latex, bend right=40] (L) to (J);
        \draw[-latex] (G) -- (H);
    \end{tikzpicture}
    \caption{Example of a directed graph with focal points and a cycle}
    \label{fig: example graph}
\end{figure}
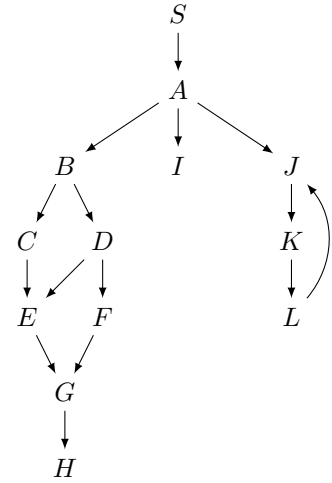

To illustrate how \cref{alg:protocol} lets cusps learn about the existence of focal points, we let the source $S$ in \cref{fig: example graph} initiate a protocol execution.
During the aggregation phase, every node queries its children for a complete list of pseudonyms for all of their children, their children, etc.
Any chain of subsequent inquiries in the left branch of the graph necessarily ends at $H$ who has no children.
The recursive chain going through $J$, then $K$, then $L$ ends when $L$ queries $J$ for the first time.
In this case, $J$ checks the Bloom filter for its secret and receives ``True'', in which case it returns its own pseudonym to $L$ without querying $K$ for the second time.
After all recursive calls have been made, pseudonyms are propagated in reverse directions, being aggregated and unblinded at each node.
Eventually, we end up with the distribution of multisets of learnt pseudonyms presented in \cref{tab: executed}.

The nodes $C$, $E$--$I$, $K$, and $L$ neither find their own pseudonym nor duplicates among their aggregated pseudonyms and can therefore conclude that they are not cusps of any focal points in the graph. Moreover, $S, A, B$, and $D$ find duplicated entries for all of their focal points and an additional duplicate for the node $H$. The latter overhead is an obvious consequence of the design of \cref{alg:protocol} since every child of a focal point which is not part of a cycle contributes at least twice to the multiset of its focal points. Still, this is in no way a contradiction to the protocol's objective.
Finally, the node $J$ has a special role in the graph as it is its own focal point which it can deduce from the presence of its own pseudonym in its multiset.

\renewcommand{\arraystretch}{1.3}
\begin{table}[h]
    \centering
    \caption{Result of \cref{alg:protocol} applied to \cref{fig: example graph}}
    \label{tab: executed}
    \begin{tabular}{|c|c|c|}
        \hline
        \textbf{Node} & \textbf{Aggregates pseudonyms of} & \textbf{Duplicates} \\
        \hline
        $S$ & $A$--$L$ & $E, G, H, J$ \\
        \hline
        $A$ & $B$--$L$ & $E, G, H, J$ \\
        \hline
        $B$ & $C$--$H$ & $E, G, H$ \\
        \hline
        $C$ & $D$, $E$--$H$ & --- \\
        \hline
        $D$ & $E$--$H$ & $G, H$ \\
        \hline
        $E, F$ & $G, H$ & --- \\
        \hline
        $G$ & $H$ & --- \\
        \hline
        $H$, $I$ & --- & --- \\
        \hline
        $J$ & $J, K, L$ & \phantom{*}---* \\
        \hline
        $K$ & $J, L$ & --- \\
        \hline
        $L$ & $J$ & --- \\
        \hline
        \multicolumn{3}{l}{\hspace{-.15cm}*$J$ detects its own pseudonym in its multiset} \\
    \end{tabular}
\end{table}

\section{Handling Compositional Risk Candidates}
\label{sec: handling}

The output of the protocol described above is the set of data processors that may be part of one or more compositional risks. However, it is not necessarily always the case that such a compositional risk really exists as these entities are just potential candidates due to their position in the data processing network. Also, the same constellation of data processors may raise a set of different compositional risks, e.g. due to different linkability issues in separate parts of the datasets processed. It hence requires additional---typically manual---inspection of each single candidate on the list to check what data is processed, how the processing works, and whether there is a compositional risk due to potential linkability between the two (or more) incoming processing requests or not, and if so, how many. 

In the context of a DPIA, the task of investigating and resolving such compositional risks lies with the data controller as being the responsible entity for the whole processing network. However, depending on the type of relationship among the other data processors in the processing network, this task may be delegated to the sub*-processor(s) in relevant position to properly handle such risks. Again, the exact entity to approach in such a case is dependent on the actual conditions of the factual data processing network, but typically, the cusps identified in the protocol would be obvious candidates to handle such compositional risks---after verifying whether an actual compositional risk exists or not.

Once a compositional risk has been identified, the data processor in charge (i.e. the cusp) may decide on the proper risk handling strategy to apply in the respective case. Generically, the data processor can decide for one of the following risk handling strategies:

\begin{description}
    \item[Accept the risk,] implying that the data processor is aware of the risk, but decides against taking any actions. This approach may e.g. be chosen if the probability of the risk manifesting or the expected damage in that case is considered to be very low.
    \item[Mitigate the risk,] e.g. by applying additional or other PETs to the data processing at appropriate positions. For the key-and-ciphertext example discussed in \cref{exampleKeyAndCiphertext}, this may e.g. imply an additional layer of encryption to be added to the secret key, the ciphertext, or both, before storing the resulting data at the cloud storage. This way, exploiting the risk requires breaking the additional encryption, which can be assumed to be non-trivial.
    \item[Eliminate the risk,] which can be achieved in different ways:
    \begin{description}
        \item[Choosing a different sub-processor] at the cusp or one of the entities involved in the composition that causes the risk is a viable strategy to eliminate the risk. For instance, in the example from  \cref{exampleKeyAndCiphertext}, a viable strategy to eliminate the risk would be to change the cloud provider utilised for storage of the secret keys at the \texttt{Database Provider}. This change of data processor would result in a different composition that does no longer pose the same compositional risk. However, due to the change of topology of the data processing network, this approach would require to re-run the protocol in order to identify and address potential new compositional risks due to the change of sub-processor.
        \item[Removing data flows] between data processors relevant for the compositional risk. This way, data that is sensitive to critical linkability would not end up at the data processors in question, hence the compositional risk is eliminated---without a change of topology. However, this strategy is not always viable due to real-world constraints on the nature of the processing.
        \item[Applying additional/other PETs] may also eliminate the risk completely. For the example in \cref{exampleCCNumber}, changing the masking approach at one of the two data processors to be consistent with the other would immediately eliminate the compositional risk, as the credit card number would then never be revealed in full at any point in the subsequent processing network.
        \item[Removing (part of) the data processing] implies to remove the respective data processors from the processing network completely, thereby eliminating the linkability risk. However, this typically also implies a reduced set of processing features that may be in conflict with business optimisation goals (e.g. with respect to data processors in the advertising industry).        
    \end{description}
\end{description}

It is worth noticing that the risk elimination strategy of \textit{choosing a different sub-processor} may cause a dilemma: If a data processor faces the risk to be excluded from the data processing network due to being part of a compositional risk, that data processor may have an incentive to hide this fact. Thus, there is an incentive for such a data processor to not properly follow the protocol described in \cref{secAlgorithm}, i.e. to deviate from the assumed honest-but-curious attacker model. For instance, such a ``malicious'' data processor may decide to remove itself from the set of risk candidates it sees, or to not even put its identifier into the list. This is why our protocol was designed with the specific goal not to reveal such a fact towards the focal points (cf. \texttt{R2} in \cref{subsec: requirements}).

\section{Discussion}\label{secDiscussion}
One of the known issues with respect to our proposed protocol and risk elicitation methodology concerns its real-world implementation. As the proposed protocol has to be implemented by several entities with different IT systems, a shared, standardised API must be provided. Looking at the developments in the European data market, the growing maturity of data spaces can be helpful here. Participants in a data space share a framework that can provide a common data format for data flows and risks upon which the protocol can be implemented. However, data-sharing scenarios are not limited to participants in the same data space. Sub-processors, such as cloud storage providers, might be used by several parties inside the data space while not belonging to it themselves. To accommodate international data sharing, the question of a platform to provide this protocol requires further attention.

Besides a shared platform to run the protocol, it must be integrated into privacy risk assessment methodologies. In a DPIA process, the protocol could be used to improve the description of data flows for a processing activity. Afterwards, the information gained by the protocol will then be leveraged for the risk assessment. To accommodate this, the DPIA process will require an exchange of information between the data controller and the data processors that includes details of the risks and measures for a data flow for each participant. 

Further, it must be examined how the proposed approach fits with privacy impact assessment regulations and methodologies worldwide. The introduced protocol leverages the hierarchy defined by the GDPR, where the data controller is responsible for the data processors involved in its processing activities. This places the data controller as the responsible entity at the top of the graph by default (node~$S$ in \cref{fig: example graph}). Different approaches to privacy impact assessments can change this ``chain of command'', or more powerful data processors in the graph might not be mandated to comply with it. 

Another issue is joint controllership (see~\cite[Art. 26]{GDPR}). The responsibility for a DPIA between joint controllers is not adequately addressed by DPIA methodologies so far~\cite{nagele_datenschutz-folgenabschatzung_2020}. However, for our methodology, it is sufficient to assume one of the joint controllers to lead the protocol execution and learn all relevant compositional risk candidates, whereas all other joint controllers behave like normal data processors according to the protocol.

Finally, it must be clarified how to deal with entities that refuse to participate in the protocol. The question of how strong the legal obligation is for a data processor to comply with the demand of a data controller at the top of the graph has to be examined in more detail. Particularly, the further you go down the tree, the relation between the data controller at the top and a data processor at the bottom of the graph gets weaker with each node.

\section{Conclusion and Future Work}\label{secConclusion}
Modern data processing instances involve collaboration of multiple data processing partners (e.g. cloud service providers) that typically use further sub*-processors, forming complex graphs of service dependencies.  
As we have shown, evaluating compositional risks to privacy in such collaboration scenarios (for example, as part of a Data Protection Impact Assessment) is non-trivial. 

We highlight the importance of this problem and argue that severe risks can arise when the same data processor is used by two (or more) different entities in the data processing network. To approach this problem, we present a protocol to detect candidates for such compositional risk manifestations in a privacy-preserving manner, i.e. without revealing the relationship between entities and their sub*-processors. We analyse the protocol's reliability and privacy properties and discuss strategies to manage the potential risks in such scenarios.

We envisage several paths for future research.
As a next step, we will implement the protocol and validate its real-world computational feasibility in simulations. We will also extend the protocol to include the risk mitigation measures. Finally, we will assess the applicability of our work to other scenarios and types of compositional risks.

\section*{Acknowledgments}
The contribution of M.J. was partially supported by Vinnova Sweden as part of the Cybercampus project.

The authors thank Tobias Pulls for his valuable input on the discussions. An AI-based tool was used to improve language and grammar of parts of the text.
\addcontentsline{toc}{section}{Acknowledgments}

\bibliographystyle{IEEEtran}
\bibliography{paper}

%\begin{thebibliography}{00}
%\end{thebibliography}

\end{document}